\documentclass[10pt,twocolumn,letterpaper]{article}

\usepackage{cvpr} 
\usepackage{times}
\usepackage{epsfig}
\usepackage{graphicx}
\usepackage{amsmath}
\usepackage{amssymb}
\usepackage{hyperref}

\begin{document}

\title{CHAMP: A Configurable, Hot-Swappable Edge Architecture for Adaptive Biometric Tasks}

\author{Joel Brogan\\
{\tt\small broganjr@ornl.gov}
\and
Matthew Yohe\\
{\tt\small yohema@ornl.gov}
\and
David Cornett\\
{\tt\small cornettdciii@ornl.gov}
\and
Oak Ridge National Laboratory\\
1 Bethel Valley Rd. Oak Ridge, Tennessee\\
}

\maketitle
\thispagestyle{empty}

\begin{abstract}
What if you could piece together your own custom biometrics and AI analysis system, a bit like LEGO\texttrademark{} blocks? We aim to bring that technology to field operators in the field who require flexible, high-performance edge AI system that can be adapted on a moment's notice. This paper introduces \textbf{CHAMP} (Configurable Hot-swappable Architecture for Machine Perception), a modular edge computing platform that allows operators to dynamically swap in specialized AI ``capability cartridges'' for tasks like face recognition, object tracking, and document analysis. CHAMP leverages low-power FPGA-based accelerators on a high-throughput bus, orchestrated by a custom operating system (VDiSK) to enable plug-and-play AI pipelines and cryptographically secured biometric datasets. In this paper we describe the CHAMP design, including its modular scaling with multiple accelerators and the VDiSK operating system for runtime reconfiguration, along with its cryptographic capabilities to keep data stored on modules safe and private. Experiments demonstrate near-linear throughput scaling from 1 to 5 neural compute accelerators, highlighting both the performance gains and saturation limits of the USB3-based bus. Finally, we discuss applications of CHAMP in field biometrics, surveillance, and disaster response, and outline future improvements in bus protocols, cartridge capabilities, and system software.
\end{abstract}

\section{Introduction}
Edge computing for performing in-situation AI tasks has become crucial in scenarios ranging from defense and security to disaster response. Field operators often face evolving mission requirements, needing to switch rapidly between AI tasks such as object detection, vehicle tracking, or biometric recognition and identification in unpredictable environments. Traditional fixed-function vision systems or cloud-based analytics are insufficient when real-time responsiveness, portability, and adaptability are required. There is a clear need for an edge AI platform that is \emph{reconfigurable in the field} by non-experts, to handle diverse tasks under power and size constraints. 

\begin{figure}
    \centering
    \includegraphics[width=0.95\linewidth]{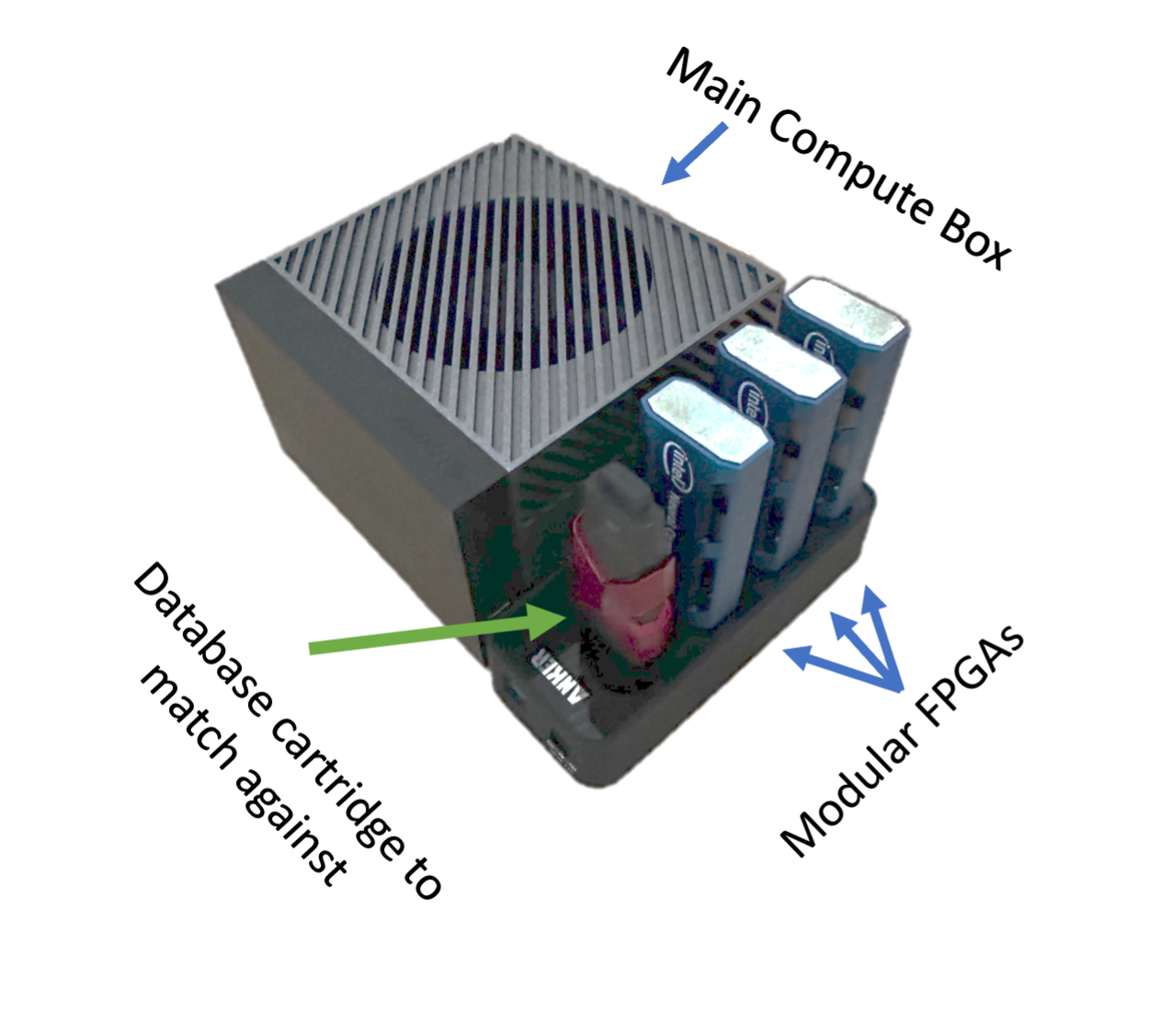}
    \caption{A depiction of the CHAMP prototype, consisting of an NVIDIA Jetson AGX Orin married with a custom high-throughput USB 3.0 bus. In this configuration, the bus is populated with with 3 Intel Movidius 2 Compute Stick modules and a final high-speed USB drive storing a large encrypted biometric database (leftmost module).}
    \label{fig:teaser}
\end{figure}

In this paper, we present \textbf{CHAMP} (\underline{C}onfigurable \underline{H}ot-swappable \underline{A}rchitecture for \underline{M}achine \underline{P}erception), an edge computing platform designed to address these needs. CHAMP provides a small, lightweight module that can be rapidly deployed and reconfigured with a variety of AI capabilities. The key innovation is a set of plug-and-play \textbf{capability cartridges}, each implementing a specific machine learning task (e.g., object detection, face recognition, natural language processing) on a low-power FPGA or neural accelerator. Operators can hot-swap cartridges to change or chain capabilities without powering down the system. The cartridges interface through a high-bandwidth \textbf{CHAMP bus} to a central controller, which runs the custom \textbf{VDiSK operating system}\cite{brogan2023vdisc} to orchestrate data flow and resource management across cartridges. Figure~\ref{fig:teaser} shows a prototype CHAMP testbed, that incorporates multiple FPGA USB accelerators on a base NVIDIA Jetson ORIN board.

The remainder of this paper is organized as follows: Section II reviews related work on modular AI systems, neural network accelerators on FPGAs, and existing orchestration software like NVIDIA's Triton. Section III details the CHAMP system design, including cartridge architecture, bus topology, FPGA integration, and the VDiSK OS. Section IV presents experimental results on performance scaling using multiple AI accelerator ``sticks'' to emulate CHAMP cartridges, demonstrating the benefits and limits of modular scaling. Section V explores use cases of CHAMP in biometrics, object tracking, document exploitation, and beyond. Section VI outlines future work to enhance the bus protocol, expand cartridge diversity, and improve the VDiSK software. Finally, Section VII concludes the paper.

\section{Related Work}
\subsection{Modular AI Systems and Hot-Swap Architectures}
The concept of modular, reconfigurable AI hardware is gaining traction as AI moves to the edge. Traditional vision systems like the Cognex In-Sight cameras~\cite{cognex_2d_machine_vision} integrate imaging and inference in a single device, but reconfiguring their algorithms typically requires external software updates and cannot be done in real-time by non-specialists. CHAMP differs by allowing physical reconfiguration of capabilities on the fly through hot-swappable cartridges, a feature more common in enterprise hardware (e.g., hot-pluggable storage or network modules) than in edge AI devices. 

On the software side, frameworks such as NVIDIA's \textbf{Triton Inference Server} support serving multiple AI models with dynamic management in cloud and edge environments.  Triton enables deployment of models from different frameworks and can load/unload models on demand to adapt to workload changes~\cite{nvidia_triton_inference_server}. This inspired aspects of CHAMP's VDiSK operating system: much like how Triton manages model ensembles and concurrency in software, VDiSK manages a collection of hardware AI modules, routing data between them and handling their addition or removal during operation. However, unlike Triton which assumes a fixed hardware platform (GPU/CPU servers), CHAMP provides a physical modularity—operators can insert a new hardware module (e.g., a face recognition cartridge) and the system will automatically incorporate it into the processing pipeline. For containers, \textbf{Kubernetes Edge}~\cite{kubernetesedge} provides centralized management and orchestration of lightweight software containers across resource-constrained edge devices. While CHAMP does not currently utilize or integrate with~\cite{kubernetesedge}, its ability to orchestrate containers over networks offers an advantage that we hope CHAMP might leverage in the future.

Prior academic work has explored modular architectures in the context of sensor networks and reconfigurable computing. For example, research in dynamic partial reconfiguration of FPGAs~\cite{lie2009dynamic} allows hardware logic to be swapped at runtime to adapt to new tasks, which parallels CHAMP's goal but at the silicon configuration level. The work in ~\cite{cornett_fpga} demonstrates reconfigurable FPGAs enable versatile, multi-purpose vision systems with competitive performance and robustness to various degradations, while significantly reducing size, weight, and power requirements for edge deployment as compared to GPU-based systems. Our approach leverages swappable hardware modules at a higher abstraction, which is more accessible to end-users than either of the previous academic implementations. CHAMP combines ideas from hot-swappable computing components and AI model management servers to create a field-deployable, user-reconfigurable AI system.

\subsection{Neural Network Accelerators on FPGAs and ASICs}
A cornerstone of CHAMP is the use of specialized accelerators in each capability cartridge. There is a rich body of work on neural network acceleration using FPGAs and ASICs, which we draw upon for cartridge design. FPGAs offer customizable parallelism and are often used to speed up deep learning inference within power or latency constraints~\cite{li2022survey}. Surveys such as Mittal's work in 2020 provide comprehensive overviews of FPGA-based CNN accelerators, highlighting techniques like model quantization and pipeline parallelism to optimize performance per watt\cite{Mittal2020}. Specific examples include Qiu \textit{et al.}'s ``DeepCNN'' accelerator on a Xilinx FPGA, which demonstrated that an embedded FPGA could run convolutional networks with significant speedups by exploiting model sparsity and 16-bit fixed-point arithmetic\cite{Qiu2016}. CHAMP's VDiSK software layer can manage interaction between arbitrary FPGA accelerators, as long as it has a software module layer that abstracts its input and output into a unified message format.  Currently, we have implemented two of these software drivers for CHAMP, specifically for the Intel Movidius Neural Compute Stick 2 (NCS2), and the Google Coral USB (GC) accelerator.

The NCS2 is powered by the Intel Movidius Myriad X Vision Processing Unit (VPU). The NCS2 is a USB-based device capable of running deep neural network inferences on only a few watts, delivering up to 4 trillion operations per second (4 TOPS) of performance~\cite{IntelNCS2}. Notably, multiple NCS2 devices can be used in parallel on one host to linearly scale throughput, as advertised by Intel\cite{IntelNCS2}. Google’s Coral Edge TPU achieves 4 TOPS on 2~W, resulting in about about 2 TOPS/W, and can run lightweight vision models like MobileNet at over 400 FPS in a power-efficient manner.

Although no further drivers have been implemented, other accelerator platforms are also good candidates for integration within the CHAMP system. ASIC accelerators like the Hailo-8 chip could push efficiency further, delivering up to 26 TOPS on ~5~W; a recent modular product by Unigen integrates two Hailo-8 chips (52~TOPS total) into an E3.S hot-swappable card for edge servers. This trend toward high-performance, low-power AI accelerators in compact form factors underlines the feasibility of CHAMP’s cartridge approach. Each CHAMP cartridge can host an FPGA or ASIC like those above, tailored to a specific AI task.

The \textbf{VDiSK operating system}~\cite{brogan2023vdisc} in CHAMP orchestrates these heterogeneous accelerators. Similar to how an OS manages different co-processors, VDiSK must handle communication and scheduling between cartridges. Prior work on multi-FPGA systems and distributed inference provides relevant insights. For instance, model-parallel approaches partition neural networks across multiple accelerators, and frameworks like OpenVINO allow deploying inference across CPU, GPU, and NCS devices concurrently. CHAMP's VDiSK builds on such ideas to dynamically allocate inference tasks to whichever cartridges are present, and ensure the data (e.g. video frames or feature tensors) flows through the chain of accelerators in the correct sequence.

\subsection{The VDiSK Operating System}
CHAMP’s runtime software, called \textbf{VDiSK} (an acronym we define as the Virtual Distributed Streaming Kernel), is a lightweight operating system designed for orchestrating modular AI pipelines. It was introduced in~\cite{brogan2023vdisc}, and has been extensively modified to provide the more complex capabilties required by CHAMP. Its role is analogous to an inference server combined with a router: it recognizes when cartridges are added or removed, queries their capabilities, and manages a message-passing interface over the CHAMP bus so that data is handed off between cartridges efficiently. CHAMP operates on a special fork of VDiSK that includes CHAMP specific features for hotswapping that we describe below. The code for CHAMP will be provided on github upon release of the paper.

\begin{figure}
    \centering
    \includegraphics[width=0.95\linewidth]{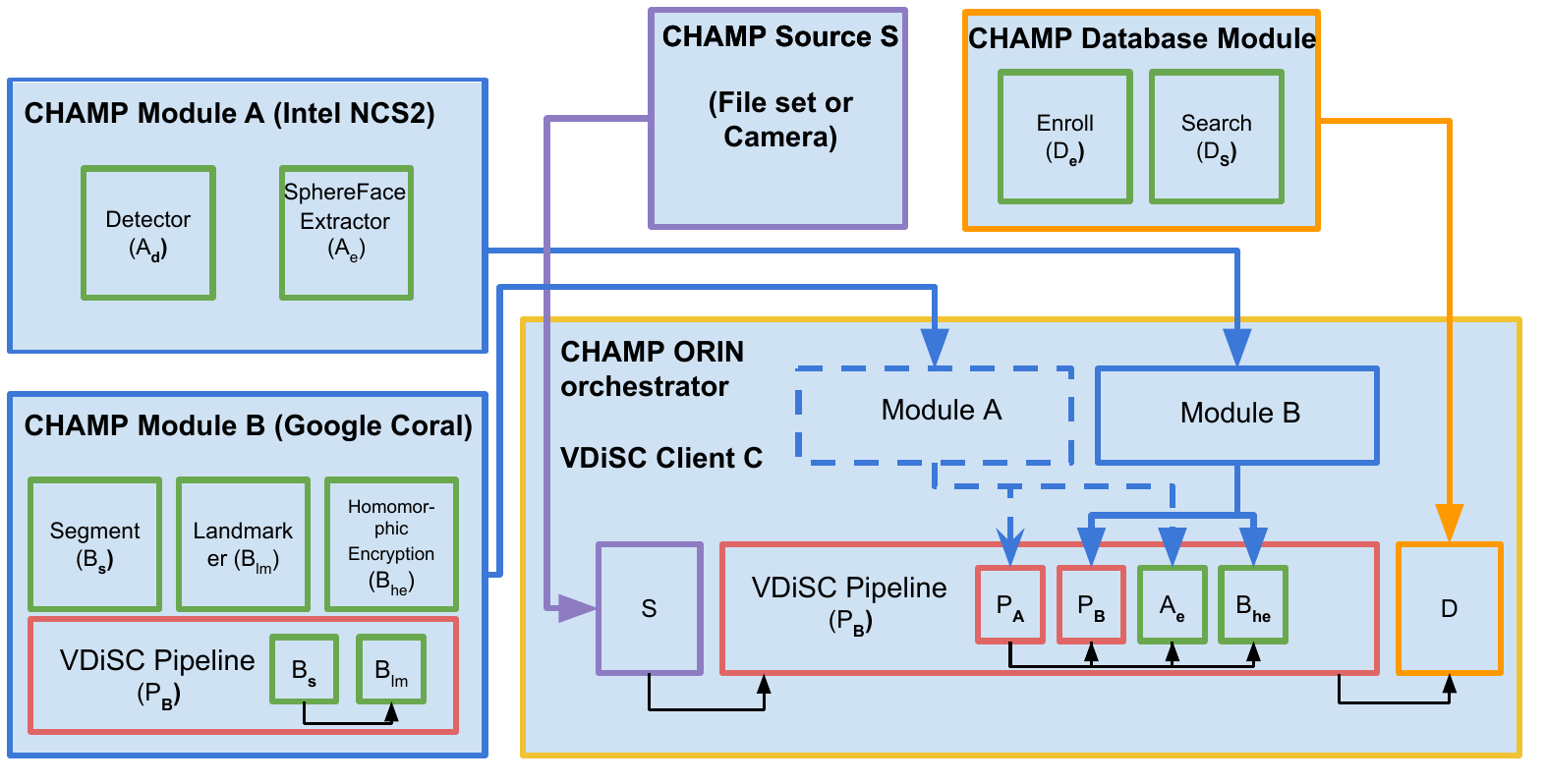}
    \caption{A visual description of the interplay between CHAMP hardware modules, the NVIDIA ORIN-based orchestrator, and the VDiSK software.  Figure is inspired by~\cite{brogan2023vdisc}.  CHAMP modules (left) provide specific capabilities that come together within the orchestrator (bottom), and can be matched against galleries within a database module (top right), which also defines the necessary matching calculation for the template type it stores.}
    \label{fig:image_set}
\end{figure}

VDiSK uses a publish/subscribe model for data exchange between cartridges, not unlike ROS (Robot Operating System)~\cite{macenski2022robot} topics for sensor data, but optimized for high-throughput streaming of imagery and vectors. Each cartridge, upon insertion, registers with VDiSK, advertising the type of data it consumes and produces (for example, “takes an image frame, outputs bounding boxes and labels”). VDiSK then links the output of one cartridge to the input of the next in a pipeline according to the physical order of cartridges or a user-specified sequence. This design was influenced by the NVIDIA Triton server's model ensemble feature, which can route outputs of one model to another internally\cite{NvidiaTriton}.

A key challenge for the CHAMP fork of VDiSK is maintaining stable operation during hot-swap events. When a cartridge is removed or inserted, the OS briefly buffers incoming data and reconfigures the pipeline routing. We ensure that if a module is removed, its upstream neighbor is instructed to pause or redirect output, and its downstream neighbor either receives a default pass-through or triggers an alert for operator intervention. Our design goal is to limit downtime to mere seconds when reconfiguring, so that the system can effectively adapt on the fly. 

The next section details the actual architecture of CHAMP's hardware and how VDiSK and the bus enable this hot-swappable capability.

\section{System Design of CHAMP}
\subsection{Overall Architecture and Bus Topology}
At the heart of CHAMP is a \textbf{Orchestrator Compute Module}, an Nvidia Jetson AGX Orin, and the \textbf{CHAMP communication bus} backplane. The bus is a off-the-shelf high-speed interface that provides both power and data connectivity to the plug-in \textbf{capability cartridges}. Physically, the bus is a multi-drop high-speed USB3.1 Gen1 bus that operates at 5 Gbps. Each cartridge connects via the standardized USB protocol. 

The bus topology allows cartridges to be arranged in a chain. Logically, cartridges form a pipeline: e.g., a first cartridge might perform object detection on video frames, passing its output (bounding boxes) to the next cartridge which performs face recognition, which then passes identified faces to a third cartridge that checks them against a biometric database, which can be encrypted via VDiSK's built-in homomorphic encryption for templates. This linear pipeline model is enforced by VDiSK, though future versions of CHAMP could allow more complex graphs (branching pipelines) with appropriate bus arbitration.

Because multiple CHAMP main modules can also be linked (for scaling out), the bus interface can be extended externally via ports that connect two CHAMP units. For example, two CHAMP modules can be connected via Gigabit Ethernet or a high-speed serial link to share data between their respective cartridge pipelines, effectively creating a larger distributed pipeline. This modular scaling means an operator could daisy-chain full CHAMP units if a task grows in complexity beyond what a single unit can handle, all while maintaining the plug-and-play ease of adding or removing pieces.

\subsection{Capability Cartridges}
Each \textbf{capability cartridge} is a self-contained AI accelerator specializing in a particular function. Internally, a cartridge consists of a low-power computational device (such as an FPGA, VPU, or ASIC), local memory, and a bus interface controller. We opted for FPGAs in our initial design due to their flexibility: a single cartridge type can be reprogrammed to a different function if needed, although in normal operation each cartridge is flashed with a fixed bitstream corresponding to its advertised capability.

The cartridges are small (a few inches in length, similar to a thick USB stick form factor) and rugged, suitable for field deployment. They are also low-power; typically each draws 5--10 Watts or less, so that the entire CHAMP system can run off battery packs if necessary to support extended field deployments. Currently implemented cartridges include:
\begin{itemize}
    \item \textbf{Object Detection Cartridge}: Cartridge that runs YOLOv3 or Mobilenet-SSD for real-time detection of people, vehicles, objects, etc. 
    \item \textbf{Face Detection Cartridge}: Implements Retinaface~\cite{deng2020retinaface} to detect facial bounding boxes
    \item \textbf{Face Recognition Cartridge}: Implements FaceNet~\cite{schroff2015facenet} and provides output embeddings to be matched against in Cosine Simlarity space.
    \item \textbf{Facial Quality Scoring Cartridge}: Implements CF-FIQA~\cite{Boutros_2023_CVPR} to provide quality socres for facial bounding boxes.
    \item \textbf{Gait Recognition Cartridge}: Implements Gaitset~\cite{chao2019gaitset} and the BodyPix~\cite{decode_bodypix} segmenting algorithm to extract embeddings from gait analysis.
    \item \textbf{Database/Storage Cartridge}: a special module that provides storage (e.g., an SSD or memory buffer) for logging data or holding large reference databases (faces) that other cartridges can query. Implements homomorphic encryption capabilities for template privacy and security from~\cite{brogan2023vdisc}.
\end{itemize}

While these cartridges only represent a small portion of possible tasks a technician may want to perform, they are a good starting set. Many more cartridges with a wide range of capabilities are in the process of being implemented, however take time to convert and distill into form factors runnable on small VPU accelerators. All cartridges conform to a common protocol for data exchange over the bus. This includes a framing for messages (e.g., image frames are tagged with sequence numbers and partitioned if large, inference results are tagged with metadata about type and size). The bus controller on each cartridge can also perform flow control; if a cartridge's processing time is slower than the input rate, it can signal upstream modules or the main controller to throttle the data flow, preventing overload.

Critically, cartridges are \textbf{hot-swappable}. The bus hardware supports live insertion: power pins are staggered so that ground makes contact first, then power, then data pins, to avoid transients. The main module monitors the bus for new connection events or removal events (using USB's standardized device detection and a Zeroconf~\cite{rfc6762}). When a new cartridge is inserted, the main module pauses the pipeline for a brief moment (a few milliseconds to a second), addresses the new cartridge, and initiates a handshake. The new cartridge reports its capability ID (a predefined code for each type of function) and its data format. VDiSK then integrates this into the pipeline at the correct position. For example, if the cartridge was inserted in slot 2 of 4, it becomes the second stage in the pipeline. Conversely, on removal, VDiSK will either bridge the gap (if the pipeline can continue without that function) or pause the pipeline and notify the operator that a capability is missing.

\subsection{VDiSK Orchestration and Software Stack}
The Main Compute Module runs a lightweight Linux with the \textbf{VDiSK OS} software stack on top. VDiSK is implemented as a set of Linux daemons and kernel modules: one kernel driver manages the physical bus (detecting devices, reading/writing from bus addresses), and a user-space daemon handles the high-level orchestration (capability registration, data routing, and health monitoring).

When the CHAMP system boots, VDiSK enumerates any cartridges present and builds an initial pipeline graph. It loads the necessary support libraries or drivers for each type of cartridge. Cartridges can operate in two modes: streaming mode (continuous data flow, e.g., video frames) or request-response mode (discrete queries, e.g., database lookup cartridge). VDiSK abstracts these differences by treating everything as a stream of messages, which could be a continuous stream or a sporadic one.

CHAMP's version of VDiSK also exposes a simple user interface for operators on a connected console or even a mobile app via WiFi/Bluetooth (as the main module has wireless connectivity. The code utilizes the ComfyUI workflow editor~\cite{comfyanonymous_comfyui} to allow an operator to see which cartridges are present and active, and can manually re-order or toggle certain pipelines via this interface if needed. However, the primary mode of configuration is physical: the operator just plugs in the cartridges in the desired order and the system auto-configures accordingly, which has been a design goal to make CHAMP usable by non-technical personnel.

\begin{figure}[ht]
    \centering
    \includegraphics[width=0.85\columnwidth]{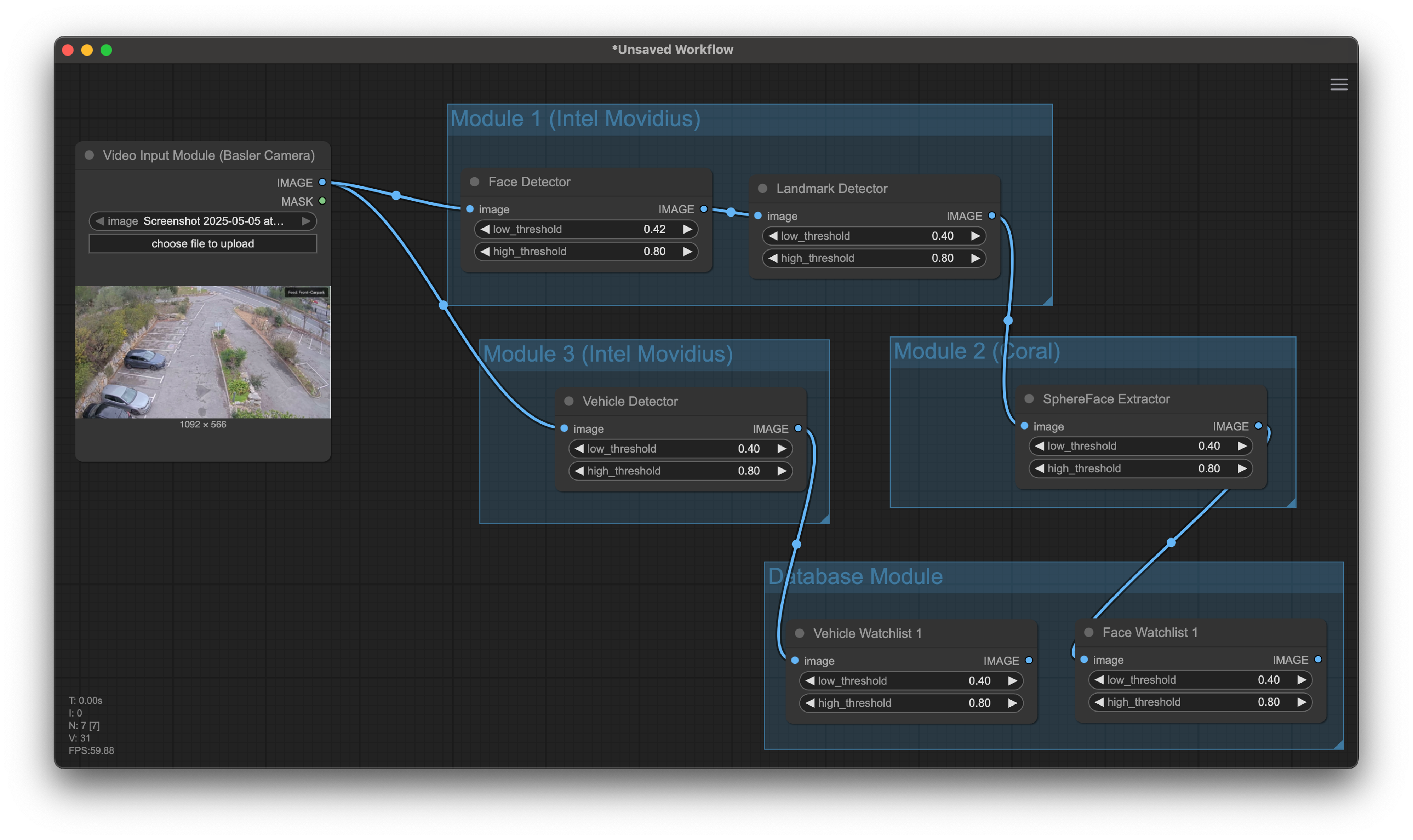}
    \caption{A sample workflow shown in the CHAMP visualization software, which is implemented within fork of the ComfyUI workflow~\cite{comfyanonymous_comfyui} editor that auto populates groups and modules based on which modules are actively plugged into the CHAMP system.}
    \label{fig:workflow}
\end{figure}

The system design emphasizes modularity and scalability: you can add capability by plugging in another cartridge, or increase throughput by linking multiple CHAMP modules. The combined hardware/software infrastructure ensures these changes happen with minimal disruption to running operations. Next, we evaluate how effective this design is via a series of experiments.

\section{Experiments and Performance Evaluation}
We built a CHAMP prototype to validate the hot-swappable design and to measure the performance scaling when multiple accelerators are used in tandem. In our current prototype, the CHAMP bus is approximated using a USB 3.0/3.1 bus with multiple \textit{Neural Compute Stick} accelerators acting as capability cartridges. While this setup differs from the final envisioned hardware (which would use a custom backplane and FPGA-based cartridges), it allows us to emulate the scenario of chaining or parallelizing inference tasks with up to five accelerators on a single bus. We focus on measuring throughput (frames per second, FPS) for a constant workload as we vary the number of active accelerator sticks, as well as observing the overhead of hot-swap operations.

\subsection{Throughput Scaling with Multiple Accelerators}
Our test scenario uses a pre-trained object detection model (MobileNetv2~\cite{sandler2018mobilenetv2}) running on Intel Movidius Myriad X sticks (Intel NCS2 devices)\cite{IntelNCS2} using the NCSDK MobileNetv2 port~\cite{movidius_ncsdk}, and the Google Coral utilizing the implementation from~\cite{tensorflow_deeplab_quantize}. The orchestrating ORIN system distributes incoming video frames to the NCS2 devices for inference. We measure the achieved frame processing rate for 1 through 5 NCS2 devices on the same USB3 bus. For fairness, we use asynchronous inferencing with a batch size of 1 per device (the model is small enough that each stick processes one frame at a time with minimal latency). We disrtribute each frame to all operating modules at once, which all perform MobileNetv2 computations simultaneously.  The experiment is performed in this way to simulate high load on the USB3.0 bus, in attempt to measure its data throughput limitations.

\begin{table}[h!]
\centering
\begin{tabular}{|c|c|c|}
\hline
\textbf{\# of Modules} & \textbf{Intel NCS2} & \textbf{Coral USB} \\
\hline
1 & 15 & 25 \\
2 & 13 & 22 \\
3 & 10 & 19 \\
4 & 8  & 17 \\
5 & 6  & 15 \\
\hline
\end{tabular}
\caption{Measured inference throughput scaling with up to five USB3 neural accelerators, each running Mobilenetv2~\cite{sandler2018mobilenetv2}. In practice, diminishing returns occur beyond 3--4 devices due to bus bandwidth and coordination overheads.}
\label{tab:scaling}
\end{table}


Table ~\ref{tab:scaling} shows the results. With a single NCS2, the system processed about 15FPS on our test video stream. We observe that as more modules are added to the system the total FPS rate decreases, likely due to high data traffic on the bus. In both experiments, FPS tapers off as more devices are added.  It should be noted that in non-simulated scenarios, each module provides a specific capability in line with the rest sequentially, so each frame is effectively pipelined through each module.  This means that as more capabilities are added, we only experience a sub-linear slowdown of operation (i.e. a system performing 500\% more compute only slows down by 50\%).

The primary cause of the slowdown is contention on the USB bus and the host CPU coordination overhead. The USB3 interface has a finite bandwidth (around 5 Gbps theoretical for USB3.1 Gen1) which in practice means the host can only send so many frames and receive results from multiple devices in parallel before hitting throughput limits. We noticed that the host CPU utilization also increased with more devices, as it has to manage multiple inference threads and USB transfers. In a future CHAMP implementation, our custom bus could mitigate some of these issues by providing higher throughput and direct module-to-module data transfers, but the experiment with USB-based accelerators is illustrative of general scaling behavior.

These results validate that adding more accelerators (capability cartridges) to a single CHAMP module can indeed increase throughput roughly linearly, until overheads set in. For many real-time applications, even 2 or 3 cartridges might suffice to meet frame-rate requirements, and CHAMP allows using just the needed number of accelerators. If further scaling is required, one could either move to a more advanced bus (e.g., PCIe Gen3/4 in a future CHAMP version) or distribute the load over multiple CHAMP modules (as noted, CHAMP modules can be networked).

\subsection{Latency and Hot-Swap Behavior}
Another set of experiments evaluated the latency added by the CHAMP pipeline and the impact of hot-swapping cartridges during operation. In a configuration with 3 NCS2 accelerators in series (simulating a 3-stage pipeline) which performed facial detection, quality estimation, and embedding extraction respectively, we measured end-to-end inference latency per frame and found it to be roughly the sum of individual device latencies plus a small overhead ($\sim$5\%) for buffer handoff between devices. This overhead is the cost of routing through VDiSK and the bus. Because CHAMP utilizes gRPC similarly to the FaRO framework~\cite{bolme2020face} and the BRIAR API~\cite{brogan2022ornl,briar-bgc2,jager2025expanding}, message passing between modules and orchestrator is extremly fast. For example, if each stick had a 30ms latency for its task, the pipeline handled a frame in about 95–100ms, indicating efficient streaming with minimal overhead.

For hot-swap, we simulated removal of the middle accelerator during runtime. Without special handling, this would break the pipeline, but VDiSK detected the removal and automatically bypassed that stage (the stage in question was a network that performs face quality estimation~\cite{EQFace}). The system paused frame processing for approximately 0.5 seconds to reconfigure, after which it resumed processing with the remaining two stages. The frames that arrived during the reconfiguration were buffered and processed afterward, meaning we did not lose data, though output was momentarily delayed. When we re-inserted the device, the system again paused for about 2 seconds to reintegrate it (slightly longer due to reloading the model on the stick), and then the 3-stage pipeline continued as before. This demonstrates the feasibility of CHAMP’s hot-swap with minimal disruption, though further refinement is needed to make it truly seamless.

\subsection{Power Efficiency}
While our current prototype did not measure power draw at a fine-grained level, we can extrapolate from known specifications of the hardware. Each Movidius NCS2 stick consumes about 1–2W when running a model continuously. Thus, five sticks might use on the order of 7–8W, and including the host overhead, the total system might be around 10~W. This is an order of magnitude lower power than a typical GPU-based inference system achieving similar throughput, underscoring the advantage of specialized low-power accelerators. One of CHAMP’s design motivations is precisely this efficiency: using many small efficient cores only when needed, rather than one power-hungry device running at all times.

\subsection{Discussion}
The experiments highlight a few important considerations for CHAMP. First, the near-linear scaling at low device counts is encouraging, as it suggests the modular approach can yield proportional benefits. Second, the saturation point we observed (~4-5 devices on USB3) will inform the choice of bus technology in CHAMP; a higher-bandwidth bus or architecture that allows parallel data paths would push this limit further out. Third, the hot-swap test, albeit simple, showed that our orchestration approach is viable. However, more complex scenarios (like swapping in a different type of cartridge that changes the data flow format) need to be tested in future prototypes. Overall, the prototype results support the core premise of CHAMP: that a configurable, hot-swappable set of AI modules can provide flexible performance on demand, with manageable overheads.

\section{Applications of CHAMP}
So who and what is CHAMP designed for? The architecture is applicable to a wide range of use cases where on-site AI processing needs to be adaptable and reliable. We outline several domains and scenarios where CHAMP can provide significant advantages:
\begin{itemize}
    \item Field Biometrics and Security: CHAMP can be deployed at checkpoints or during tactical operations to perform face recognition, fingerprint matching (with new implemented modules), or gait recognition for individuals in real-time. An operator can carry a CHAMP unit with cartridges for face detection and face ID matching against a watchlist. If the mission changes (e.g., now needing vehicle recognition or thermal imaging at night), the operator can swap in the appropriate cartridges (vehicle detector, thermal camera module) on the fly. This is highly relevant for military Special Forces or law enforcement units that encounter rapidly evolving scenarios.
    \item Object Detection and Tracking: For surveillance or search-and-rescue, CHAMP can serve as a portable analytics hub processing video streams from drones or cameras. One configuration might include a wide-area motion detector cartridge, a target classification cartridge, and a tracker cartridge. In disaster response, an operator could use CHAMP with a drone feed: initially slot in a debris detection module to identify blocked roads, then replace it with a human body detection module to look for survivors, depending on mission needs.
    \item Disaster Response and Environmental Monitoring: In a natural disaster, communication networks might be down, so edge processing is crucial. CHAMP modules could be set up to analyze sensor data for signs of life (image, sound sensors), monitor structural damage via vision algorithms, or even run predictive models for aftershocks or weather patterns if needed. The modularity allows responders to tailor the analytics to the situation at hand by choosing the right set of cartridges.
    \item Mobile Device Forensics: Law enforcement can use CHAMP to perform on-site analysis of mobile phones or laptops obtained in the field. For instance, a CHAMP unit could include a data extraction cartridge and a machine learning cartridge to flag illicit content (like an image classifier for contraband or a clustering algorithm to visualize communications). The advantage is that this analysis can happen on-scene (e.g., at a border crossing or remote location) without relying on cloud connectivity, and if new types of analyses are needed, a new cartridge (say, a deepfake detection module for videos) can be inserted into the existing setup.
    \item Industrial and Commercial Vision Systems: Outside of government uses, CHAMP could be deployed in factories for quality control. Unlike fixed vision systems, a CHAMP-based solution would let technicians swap in new defect detection modules as products change, without overhauling the entire system. Similarly in retail, a CHAMP unit could be used for different analytics (customer counting, shelf inventory checking, etc.) by switching cartridges based on store needs.
\end{itemize}

These examples demonstrate CHAMP’s versatility. The common theme is that CHAMP brings a \emph{LEGO\texttrademark{}-like} plug-and-play experience to AI deployment: just as one can build different structures with the same set of LEGO\texttrademark{} bricks, one can assemble different AI pipelines with a set of CHAMP cartridges. Moreover, because it is edge-based, it supports real-time decision making in the field and preserves data privacy (no need to stream sensitive data to cloud).

In particular, defense and intelligence applications stand to benefit greatly, as noted by potential users in DoD, DHS, FBI contexts. The ability to carry a small box into a mission that can be quickly reconfigured for whatever analytic task is needed (without having to pre-define that at deployment time) could enhance situational awareness and operational flexibility.

\section{Future Work}
While our current CHAMP prototype and design show promise, there are several areas for significant further development to fully realize the vision of a robust, flexible platform. The present design uses an off-the-shelf USB3-based bus for expedience, but a custom bus that allows for less memory transfer overhead, and higher throughput is in development. Future work will explore using USB-C, PCIe or even proprietary serial links that offer higher throughput and lower latency. We aim to design a bus protocol that supports direct peer-to-peer cartridge communication (so intermediate data can go from one cartridge to the next without always traveling up to the main controller). We will also investigate dynamic reconfiguration of the bus to allow cartridges to be re-routed in different topologies (not just a fixed pipeline). This type of operational mode could be useful in multi-camera or multi-sensor systems that require parallel coordination and fusion of sensor output, such as multi-camera neural high-dynamic-range post processing techniques~\cite{10.1117/12.2566765}.

As CHAMP is further developed, we will perform more in-depth experiments thoroughly analyzing the latency, performance, and power consumption dynamics of CHAMP under different, more intricate biometric workloads. Furthermore, as the template privacy capability within VDiSK is refined, we will perform exhuastive experiments on the speed and power requirements of running privacy-preserving template encryption and matching techniques in-line with these workloads, as this type of privacy protection is an important component of running biometric systems at the edge.

Additionally, we plan to diversify the types of cartridges available. Developing new cartridges will also involve porting state-of-the-art models to our low-power hardware. We hope to leverage techniques like model pruning, quantization to low-bit (e.g., 4-bit or binary neural networks), and even model distillation to fit big AI capabilities into small cartridges. As we develop CHAMP, we see the potential for a broader ecosystem. We aim to standardize the cartridge interface protocol so that third-party developers could create new cartridges and drivers compatible with CHAMP. We would like to extend CHAMP to also work with accelerators such as the EdgeCortix SAKURA-II~\cite{sakuraii_ec2025} and Hailo-8~\cite{hailo8_2025}

Finally, an exciting area of research is how CHAMP can facilitate multi-modal AI. Because it can host different types of processing units simultaneously, we plan to experiment with pipelines that fuse, for example, image and audio data for better scene understanding and biometric matching. With appropriate synchronization support in VDiSK, one could have a microphone cartridge and a camera cartridge both feed into a fusion module that does audio-visual processing (e.g., detect a person and their speech together to do speaker identification). The flexibility of CHAMP could make setting up such multi-modal pipelines much easier than custom-building a single device for each combination of sensors.

\section{Conclusion}
We have presented CHAMP, a novel configurable and hot-swappable architecture tailored for real-time machine perception tasks at the edge. By combining modular hardware AI cartridges with a high-speed bus and an intelligent orchestration OS (VDiSK), CHAMP enables on-demand reconfiguration of AI capabilities in the field. This approach empowers non-technical users to adapt their AI toolset to changing missions simply by swapping modules, rather than needing to carry multiple devices or perform software reprogramming under pressure.

Our design and prototype demonstrate that CHAMP can scale performance by adding accelerators, achieving up to fourfold increases in throughput with five modules, and that it can maintain operation with minimal downtime during module swaps. These features differentiate CHAMP from static edge AI appliances and make it particularly suited for applications in defense, security, emergency response, and flexible industrial automation. Whether it’s identifying persons of interest, translating foreign documents on the fly, or rapidly deploying a new sensor analysis capability during a disaster, CHAMP provides a foundation for agile AI deployment.

Moving forward, we are focused on refining the bus and OS for greater performance and robustness, expanding the library of plug-and-play AI cartridges, and engaging with end-users to drive CHAMP towards real-world deployments. As edge computing continues to grow in importance, we believe modular architecture paradigms like CHAMP will play a key role in delivering AI whenever and wherever it’s needed, with versatility to meet the unknown challenges of in-situation fluidity.

\section{Acknowledgements}
Notice:  This manuscript has been authored by UT-Battelle, LLC, under contract DE-AC05-00OR22725 with the US Department of Energy (DOE). The US government retains and the publisher, by accepting the article for publication, acknowledges that the US government retains a nonexclusive, paid-up, irrevocable, worldwide license to publish or reproduce the published form of this manuscript, or allow others to do so, for US government purposes. DOE will provide public access to these results of federally sponsored research in accordance with the DOE Public Access Plan (http://energy.gov/downloads/doe-public-access-plan).

Research sponsored by the Laboratory Directed Research and Development Program of Oak Ridge National Laboratory, managed by UT-Battelle, LLC, for the U. S. Department of Energy.
{\small
\bibliographystyle{ieee}
\bibliography{egbib}
}

\end{document}